\documentclass[sigconf]{acmart}

\usepackage[font=footnotesize,labelformat=simple]{subcaption}

\copyrightyear{2025}
\acmYear{2025}
\setcopyright{cc}
\setcctype{by-sa}
\acmConference[CI 2025]{Collective Intelligence Conference}{August 4--6, 2025}{San Diego, CA, USA}
\acmBooktitle{Collective Intelligence Conference (CI 2025), August 4--6, 2025, San Diego, CA, USA}
\acmDOI{10.1145/3715928.3737487}
\acmISBN{979-8-4007-1489-4/2025/08}

\begin{document}

\title{Case Law Grounding: Using Precedents to Align Decision-Making for Humans and AI}

\author{Quan Ze Chen}
\email{cqz@cs.washington.edu}
\affiliation{%
  \institution{University of Washington}
  \city{Seattle, WA}
  \country{USA}}

\author{Amy X. Zhang}
\email{axz@cs.washington.edu}
\affiliation{%
  \institution{University of Washington}
  \city{Seattle, WA}
  \country{USA}}


\newcommand{\bug}
    {\mbox{\rule{2mm}{2mm}}}
\newcommand{\chm}[1]{#1}

\begin{abstract}
From moderating content within an online community to producing socially-appropriate generative outputs, decision-making tasks---conducted by either humans or AI---often depend on subjective or socially-established criteria.
To ensure such decisions are consistent, prevailing processes primarily make use of high-level rules and guidelines to ground decisions, similar to applying ``constitutions'' in the legal context.
However, inconsistencies in specifying and interpreting constitutional grounding can lead to undesirable and even incorrect decisions being made.
In this work, we introduce ``case law grounding'' (CLG)---an approach for grounding subjective decision-making using past decisions,  similar to how precedents are used in case law.
We present how this grounding approach can be implemented in both human and AI decision-making contexts, introducing both a human-led process and a large language model (LLM) prompting setup.
Evaluating with five groups and communities across two decision-making task domains, we find that decisions produced with CLG were significantly more accurately aligned to ground truth in 4 out of 5 groups, achieving a 16.0--23.3 \%-points higher accuracy in the human process, and 20.8--32.9 \%-points higher with LLMs.
We also examined the impact of different configurations with the retrieval window size and binding nature of decisions and find that binding decisions and larger retrieval windows were beneficial.
Finally, we discuss the broader implications of using CLG to augment existing constitutional grounding when it comes to aligning human and AI decisions.
\end{abstract}





\maketitle

\section{Introduction}

Many technological systems today are entangled in \textit{social} decision-making situations. 
Online spaces for inter-personal communication, like social media platforms, need to contend with decisions around how to moderate content in a way that maintains a healthy community~\cite{gongane2022detection}. 
General purpose AI assistants, like LLM-backed chatbots, also need to implicitly make decisions around how to generate answers that users find appropriate~\cite{ji2023ai,shen2023large,gabriel2020artificial}.
The scale of these systems means that getting the decisions wrong can have major negative consequences for people~\cite{haimson2021disproportionate,Weidinger2021EthicalAS}.

Unlike systems that work with more objective determinations, decisions around social concepts are especially hard because: (1) decision criteria can vary across different groups and tasks; and (2) it can be difficult (or even impossible) to formalize these criteria accurately.
Efforts to produce high-quality decisions often involve significant effort, making use of deliberation~\cite{im2018deliberation,small2021polis}, juries of community members~\cite{fan2020digital}, and even expert panels~\cite{klonick2020facebook}.
Scale also creates additional challenges, ultimately leading to the need for automated tools~\cite{Jhaver2019HumanMachineCF,wulczyn2017ex} or paid crowd workers~\cite{satariano2021silent} to make final decisions, both of which have introduced inconsistencies and errors.

In order to improve the consistency of decisions, processes often first create artifacts to \textit{ground} the decision-making, the most common of which takes the form of rules and guidelines (which we collectively refer to as \textit{constitutions}).
For example, moderators of communities on social media platforms often create community guidelines that establish what content is allowable, and how violations are to be dealt with~\cite{Chandrasekharan2018redditnorms,chandrasekharan2019hybrid}.
Similarly, builders and operators of AI assistant services often come up with principles~\cite{huang2024collective} that are then used to synthesize training and fine-tuning examples~\cite{bai2022constitutional} or built into system prompts~\cite{lin2023unlocking}.
However, rules and criteria make use of imprecise language that can leave room for ambiguity~\cite{fiesler2018reddit,chen2021goldilocks} around edge cases, and varied decision-makers often have divergent interpretations~\cite{gordon2021disagreement}.
Is there another way we could better support these subjective decisions at scale?

Taking inspiration from case law in legal systems, we introduce \textbf{case law grounding} (CLG), a new approach to grounding decision-making processes by making use of past decisions (\textit{precedents}) rather than rules (\textit{constitutions}) as guidance.
CLG accomplishes this through 3 steps: 
(1) case \textbf{retrieval}, where past cases potentially relevant to the current decision are retrieved; 
(2) precedent \textbf{selection}, where retrieved precedents are further \textit{refined} for applicability to the current decision; and finally 
(3) decision synthesis, where the selected precedents and decisions are aggregated and synthesized into the decision for the new case.
To address scaling needs, we implemented this grounding approach both as a \textit{human-led} version where we support human decision-makers (e.g., members of a community or group), and as a \textit{LLM-prompting} version where an LLM-backed agent is making the decisions.

We evaluated our system on two decision-making tasks: content \texttt{mod}eration, where binary ``keep''/``remove'' decisions were made on Reddit comments~\cite{park-etal-2021-detecting-community}, and \texttt{tox}icity rating, where social media posts were rated on a 5-point scale~\cite{kumar2021designing}.
Recruiting participants from 3 Reddit communities ($n=35$) and 2 demographically-determined population groups ($n=60$), we evaluated the human-led version of CLG against a baseline approach that used established constitutions (community guidelines and toxicity annotation guidelines) to support grounding.
Using off-the-shelf LLMs, we also evaluated the \textit{LLM-prompting} version of CLG against a similar baseline where LLMs conducted decision-making using the same established guidelines.

In summary, we make the following contributions:
\begin{itemize}
    \item We introduce a general framework for supporting decision-making in subjective and social contexts, \textbf{case law grounding} (CLG), and implement two versions: a \textit{human-led} process, and an \textit{LLM-prompting} process.
    \item We evaluate CLG across two decision task domains (spanning a total of 5 different group-aligned decision tasks), comparing CLG against a more traditional approach that uses constitutional grounding. We find that:
    \begin{itemize}
        \item In both the human-led or LLM-prompting process, CLG was able to produce final decisions that more accurately matched observed ground truth compared to decisions grounded with constitutions, with statistically significant improvements observed in 4 out of 5 group decision tasks. (RQ1)
        \item For most domains, larger retrieval windows resulted in higher accuracy, however scaling characteristics varied across groups. (RQ2a)
        \item Using \textit{binding} precedents that directly affect the final decision is important---with the 4 groups showing statistically significant improvements only doing so in the binding version. (RQ2b)
    \end{itemize}
    \item We discuss our findings in the broader context of grounding subjective, socially-situated decision-making tasks, noting the benefits and drawbacks of each type of grounding, and point to how \textit{case law} and \textit{constitutional} grounding can complement each other in supporting group-aligned subjective decision-making.
\end{itemize}
\begin{figure*}
  \centering
  \includegraphics[width=0.9\linewidth]{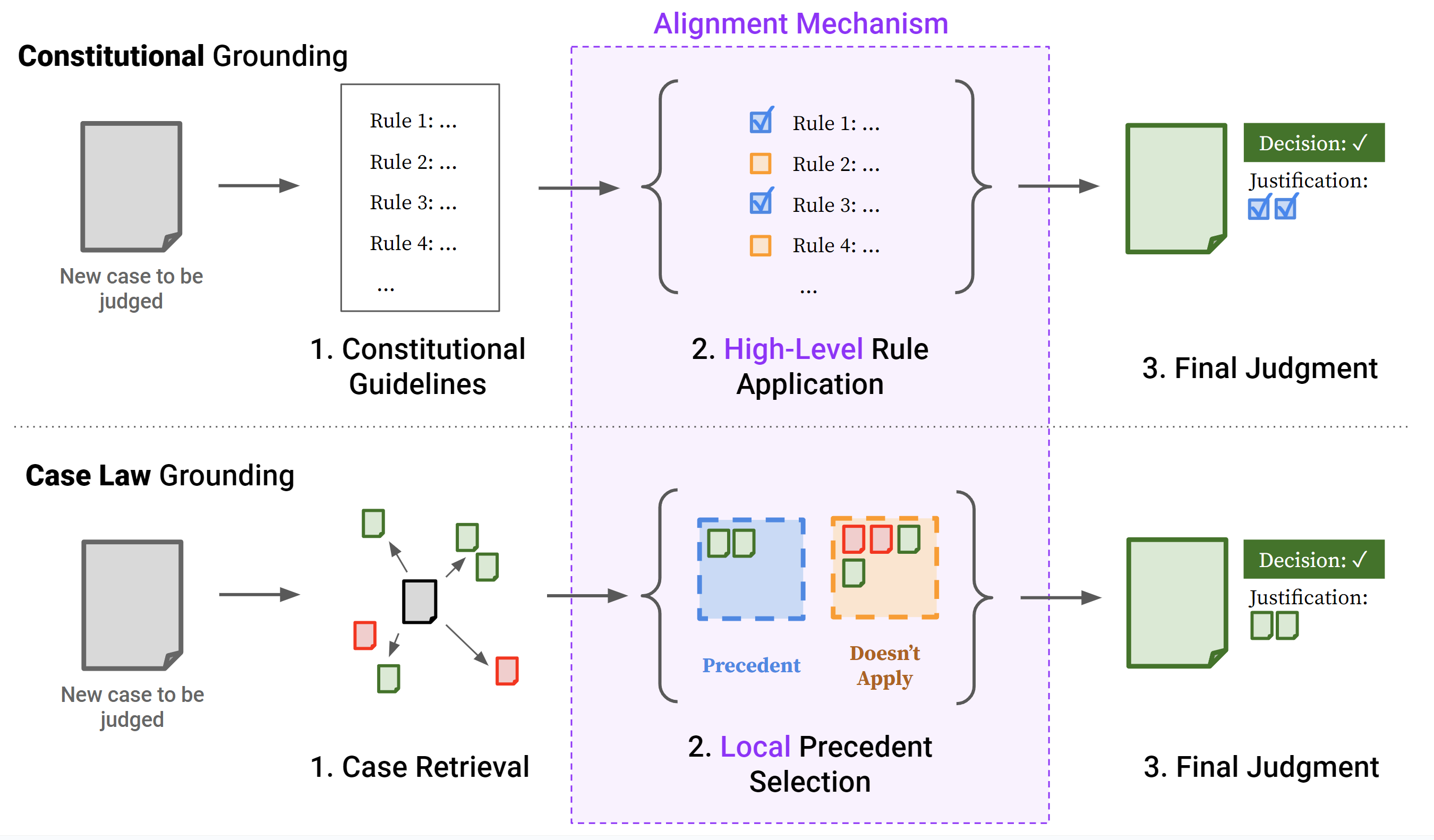}
  \caption{A diagram contrasting different approaches to align decisions: (1) traditional ``constitutional'' grounding design, involving applying abstract guidelines to construct judgments; and (2) ``case law'' grounding proposed in this work, that utilizes retrieval of past cases followed by selection of applicable precedents to construct judgments.}
  \label{fig:diagram:alignment-mechanisms}
  \Description{A diagram illustrating two types of ways to ground decisions. On the top is a constitutional workflow, which follows a process of referring to guidelines, applying high-level rules, and then making a final judgment. Below is a case law workflow, which follows a process of retrieving relevant cases, selecting locally-relevant precedents, and then automatically deriving a final judgment. The second step in each process is highlighted purple indicating it forms the core of what is an alignment mechanism.}
\end{figure*}

\section{Related Work}

In this section, we discuss how our work relates to prior work by examining the challenges around subjective and social decision-making, designs for and applications of constitutional grounding, and precedents for cases to aid decision-making in prior social and technical systems.  

\subsection{Challenges in Grounding Subjective and Socially-Established Concepts}

Subjective and socially-defined concepts are common in many situations where humans or AI systems need to make decisions, posing unique challenges. 
For example, online communities on social media platforms often have group-specific shared knowledge in the form of norms or ``common sense'' criteria around what content (or behavior) is appropriate~\cite{Chandrasekharan2018redditnorms}.
These norms often contribute significantly towards how moderation decisions are made~\cite{matias2019civic}, but frequently are not explicitly documented~\cite{juneja2020through}.
This often makes it challenging for those without such context to accurately judge situations even in the presence of written community guidelines~\cite{jhaver2019-reactions,jhaver2019does}.
Additionally, concepts in written guidelines can often be under-specified or even conflicting, which can lead to diverging interpretations~\cite{weld2021making,gordon2021disagreement}.
Furthermore, socially-established norms and criteria are often fluid, changing over time as the community acquires new members or demographics shift~\cite{lin2017better}.
These factors all make it hard to create and maintain artifacts for ``grounding'' these concepts such that they are understood consistently across people and over time. 

Beyond online communities, even annotation tasks can involve some degree of subjectivity.
For example, instructions or task designers that seem clear to crowdsourcing requesters may actually be confusing to members of a crowd~\cite{Wu2017ConfusingTC}.
Surveys or rating tasks can also involve questions that are by nature subjective~\cite{bentley2017comparing}, ambiguous, or a combination of both~\cite{Chen2023JudgmentSR}.
Efforts in the crowdsourcing realm to build more effective grounding often end up running into conflicting goals---where concepts need to be both unambiguously specified so the annotation can be done while also leaving just the right amount of flexibility for providing the exact subjectivity that we want to capture.

\subsection{Rules, Guidelines, and Constitutions}

On big social network platforms that host content for millions or billions of people, consistency and transparency around moderation play a paramount role in user trust, making it important to have a clear set of content guidelines and even training materials that ground moderation decision-making~\cite{times,klonick2020facebook}.
These sets of constitutional rules are costly to develop and often require training of moderators to execute consistently---largely relegating them to bigger organizations where such costs can be justified~\cite{roberts2014behind,satariano2021silent}
Even in smaller online communities, rules and community guidelines often serve as ways to ground the criteria for what is acceptable content~\cite{fiesler2018reddit}. 
However, prior work has also observed that in many cases real-world moderation decisions do not always match the expectations of users even given community guidelines~\cite{haimson2021disproportionate,jhaver2019-reactions,jhaver2019does}.

Indeed, not all constitutional grounding needs to take the form of rules and guidelines applied by humans and thus subject to subjective interpretation.
Online communities have built constitutions in the form of code to automatically execute decisions, such as with systems like AutoModerator~\cite{Jhaver2019HumanMachineCF}, ORES~\cite{Halfaker2019ORESLB}, and PolicyKit~\cite{Zhang2020PolicyKitBG}.
Constitutional grounding can also form grounding that is internally applied to ground model behavior, such as with efforts around Constitutional AI~\cite{bai2022constitutional,huang2024collective} where humans provide guidelines that ground the behavior of LLMs.
However, even in the case of constitutions implemented as code, models can introduce new biases, and the problem of incomplete or conflicting constitutions can still cause decisions that do not align with the intentions of groups or communities.

\subsection{The Case for Cases}

Cases and case studies have broadly been used as a medium for reasoning around complex scenarios in various fields similar to the socially-constructed judgments we focus on. 
For example, moral case deliberation~\cite{pierce2013morality,sush2021workbook,emelin-etal-2021-moral} is often used in medical settings to reflect on ethical situations~\cite{Molewijk2008TeachingEI}, and case studies are common in business educational settings too~\cite{Farashahi2018EffectivenessOT}.
In legal education, fictional cases are frequently used in exercises as a medium for engaging in deliberative reasoning~\cite{gaubatz1981moot,ringel2004designing}. 
Case-based reasoning (CBR) that focuses on analyzing the circumstances around specific case scenarios also has a long and rich history as a method for thinking about moral philosophy~\cite{paulo2015casuistry} and in the realm of jurisprudence in the US legal system~\cite{adversarial,Caputo2024AlignmentAJ}.

In crowdsourced annotation, cases in the form of examples can augment rubrics to demonstrate how judgments should be made~\cite{Wu2017ConfusingTC}, act as tests to evaluate whether a concept or criteria was accurately understood~\cite{liu2016effective,hettiachchi2021challenge,singla2014near}, and even pose as reference anchors to ground the levels on a subjective scale~\cite{chen2021goldilocks}.
More broadly speaking, a data-centric view around subjective concepts posits that data points can also be broadly interpreted as cases~\cite{simons2020hope,jiang2022investigating}, with the training of classifiers using human-annotated data being itself a form of grounding (a model's) decisions through the medium of cases.
Though in many datasets, issues around the original data collection often limit their utility as a good model of subjectivity~\cite{Arhin2021GroundTruthWT,denton2021whose}.

Finally, the detail provided by individual cases means that they can act as a medium for deliberation to reconcile differences and work towards shared understanding.
We can see examples of this in the legal realm around adjudication~\cite{carpenter1917court}, medical realm for diagnosis~\cite{bertens2013use,schaekermann2019understanding}, in communities for working out moderation policy~\cite{fan2020digital,pan2022-legitimacy-digital-juries}, and more generally just around challenging tasks involving different lines of reasoning~\cite{chen2019cicero}.
\section{Case Law Grounding}
\label{sec:design}

In the legal space, there are similar challenges when it comes to grounding decisions in a way that follows the will of the public, and prior work has observed similarities between the grounding of decisions in sociotechnical settings and those in the legal realm~\cite{Caputo2024AlignmentAJ}.
One interesting parallel lies in the mechanisms used for grounding decisions: similar to the rules and guidelines common in sociotechnical systems, constitutional (or statutory) laws also serve a similar purpose of providing high-level guidance around decisions.
However, in addition to constitutional grounding, some legal systems also make use of \textit{case law} to address situations when constitutional guidance is ambiguous, incomplete, or conflicting.
The concept of case law presents the idea that past decisions can also ground future decisions in the form of analogies between past cases and the current case. 
By identifying similarities and differences in the nature of facts within each case, it is possible to derive ad-hoc unstated principles or determinations that people are in agreement on~\cite{ashley1989modelling}. 

In this work, we propose Case Law Grounding (CLG) as an initial approach that explores the idea of using past decisions to ground decision-making (\autoref{fig:diagram:alignment-mechanisms}). In the following subsections, we will describe a general set of steps for conducting case law grounding in the decision-making process, followed by two specific implementations of this process in a \textit{human-led} and an \textit{LLM-prompted} design.

\begin{figure}[h]
  \centering
  \includegraphics[width=\linewidth]{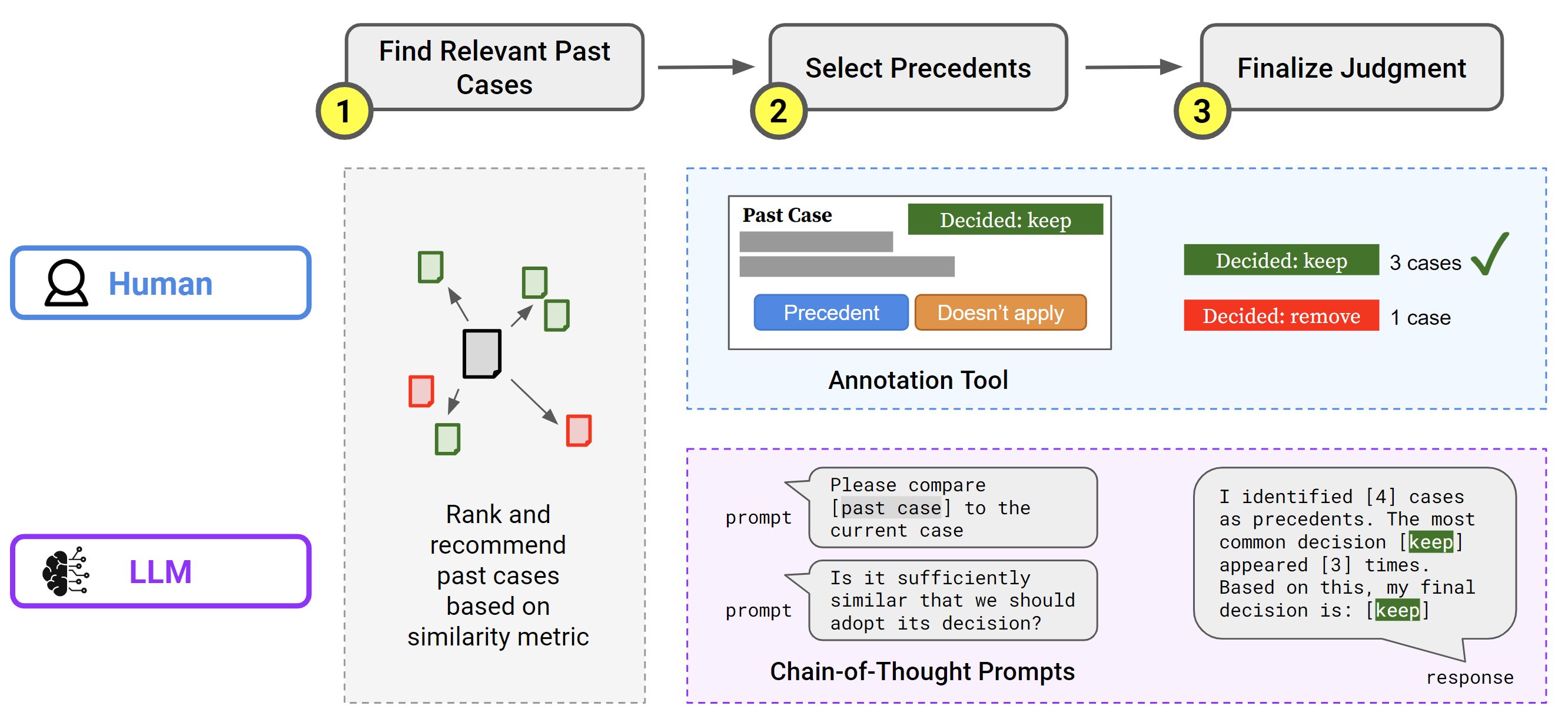}
  \caption{A diagram showing, at a high level, the process of \textbf{case law grounding} reflected as 3 steps: (1) finding potentially relevant past cases, (2) selecting precedents, and (3) finalizing precedents into judgments.
  We also show two instantiations of the process: A \textbf{human-centric} process (top) where, after relevant cases are retrieved, human annotators use an interactive tool to view and classify cases, and an \textbf{AI-centric} process (bottom) where an LLM takes the role of a decision-maker and, through chain-of-thought style prompts, evaluate precedents and form decisions.}
  \Description{A diagram of the case law grounding process implemented}
\end{figure}

\subsection{Step 1: Retrieving Relevant Cases}
\label{sec:design:step:1:case-retrieval}

The first step in case law involves conducting legal research. 
In legal practice, lawyers spend considerable effort to locate, examine, and reason about the space of past decided cases in order to draft the best argument for their client. 
This often involves querying legal databases to identify relevant cases that can be \textit{binding} precedents for the case they are arguing~\cite{berring1987legal}. 
Once relevant cases are found, there can be additional criteria and checks to make sure that they can apply.

\textbf{Semantic Similarity for Case Retrieval}:
Experts in the legal space depend on specialized keywords and indexing systems that let them navigate well-reasoned \textit{facets} of case law to quickly localize their search space.
These indices are valuable because they can encode crucial aspects of a case that directly help identify its viability as a precedent (e.g., is the jurisdiction the same, are there commonalities in the circumstances, evidence, or statutory nature of the case); thus it makes sense to spend resources to create them.
In the settings we consider, we can't expect to have ready-made case indices, but we can utilize advances in general-purpose language models to approximate case retrieval through text embeddings.
Since the cases we're interested in (e.g., online comments, interpersonal advice scenarios, user queries, and AI model responses) encode all of their factual properties through their natural language descriptions, a reasonable proxy for the relevance of any case to any other can be approximated in the form of the distance between them in the embedding space $d(x_\text{past}, x_\text{current})$. Depending on the domain, it may be possible to choose an embedding specifically created for the domain, though in many situations, an off-the-shelf generic embedding can also be sufficient.

\subsection{Step 2: Determining Applicable Precedents}
\label{sec:design:step:2:determine-precedents}

Simply surfacing potentially relevant cases to the agent making a decision is likely not sufficient, as we don't know whether the decision-maker considers the prior case as precedent setting and whether they have integrated it into their judgment.
In the legal realm, the selection of precedents from related cases is made explicit, taking the form of a written document containing the arguments being made, and especially the citations to prior precedent cases that support such arguments.
Indeed, this explicit identification and documentation of precedents as supporting arguments to a judgment can confer many benefits. For one, by documenting the exact cases used to support a judgment, it becomes easier to audit the reasoning behind judgments. 
One can gain an understanding of disagreements between different decision-makers by inspecting the commonalities and differences in the precedents selected. 
Additionally, selecting precedents decouples the judgment process from a more subjective (and possibly un-grounded) \textit{interpretation} of a case and outcome to instead focus on \textit{reasoning} about more objective comparisons between aspects of the cases. 
Finally, by having a shared set of precedents to draw from, we can improve the consistency of decisions across different adjudicators who may have different personal subjective preferences.

However, requiring a full argument for each case, while effective for challenging crowd judgment tasks~\cite{drapeau2016microtalk,chen2019cicero}, can be labor intensive.
To mitigate this issue, in this stage we simplify the argument process to focus only on the classification of whether a retrieved case is a good precedent, not why it is so. This means that instead of coming up with a judgment, writing a justification, and citing cases to support it, an adjudicating agent focuses on presenting their evaluation of the neighborhood of cases.

Specifically, in this step, the agent will be asked to classify the retrieved cases into either:

\begin{itemize}
    \item \textbf{Precedent\footnote{Often referred to as an ``adopted'' precedent.}}: A retrieved case should be marked as a precedent if that prior case provides support for the judged case to adopt a similar decision due to substantially similar circumstances or aspects (i.e., cases that fall on the same side of the decision bound of the judged case).
    \item \textbf{Doesn't Apply\footnote{Often referred to as a ``distinguished'' precedent.}}: A retrieved case does not apply as a precedent if there are critical aspects of the case that differ from the judged case such that the argument for the past decision would not apply to the new case.
    While many cases that don't apply as precedents are likely to fall on the opposite side of the decision bound of the judged case, this is not a requirement for the case to not apply.  
    Sufficiently significant \textit{differences} in circumstance or factual aspects don't necessarily need to result in \textit{opposing} outcomes (e.g., involvement of a minor versus adult in a legal case might affect the argument but the outcome judgment could end up being the same).
\end{itemize}

While case selection can be led by a human annotator, with advances in reasoning with natural language using language models, it is also possible for an LLM to be used with a specially designed prompt to complete this step.

\subsection{Step 3: Deriving Judgments from Selected Precedents}
\label{sec:design:step:3:deriving-judgments}

Once the sets of precedent cases have been selected in the step above, a final decision can be synthesized by checking each case in the precedent sets and adopting the judgments associated. 
With the assumption of \textit{stare decisis}, we view the previous judgments on any retrieved past cases as a reliable source of truth. 
This means we can view each case as evidencing a direction to take the final judgment.

\subsection{Supporting Human-Led and LLM Prompted Decision-Making}
\label{sec:design:workflows}

\begin{figure*}[ht]
  \begin{subfigure}[t]{0.6\linewidth}
      \vskip 0pt  
      \includegraphics[width=\linewidth]{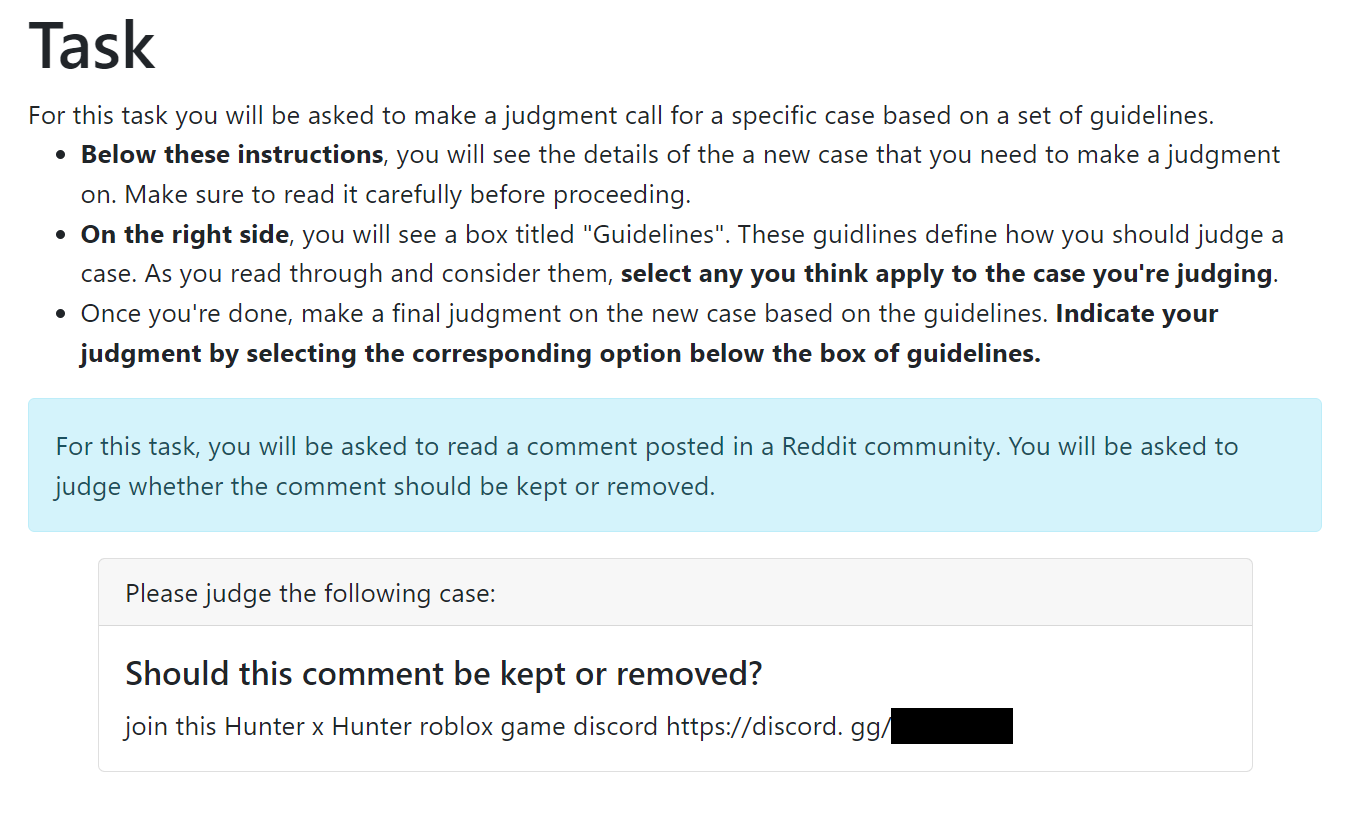}
      \caption{Task and instructions}
      \label{fig:sc:user-interface:instructions}
      \Description{Screenshot of the task and instructions of the interface, with general instructions on how to use the annotation tool at the top as a bulleted list,}
  \end{subfigure}
  \hfill
  \begin{subfigure}[t]{0.48\linewidth}
      \vskip 0pt
      \includegraphics[width=0.9\linewidth]{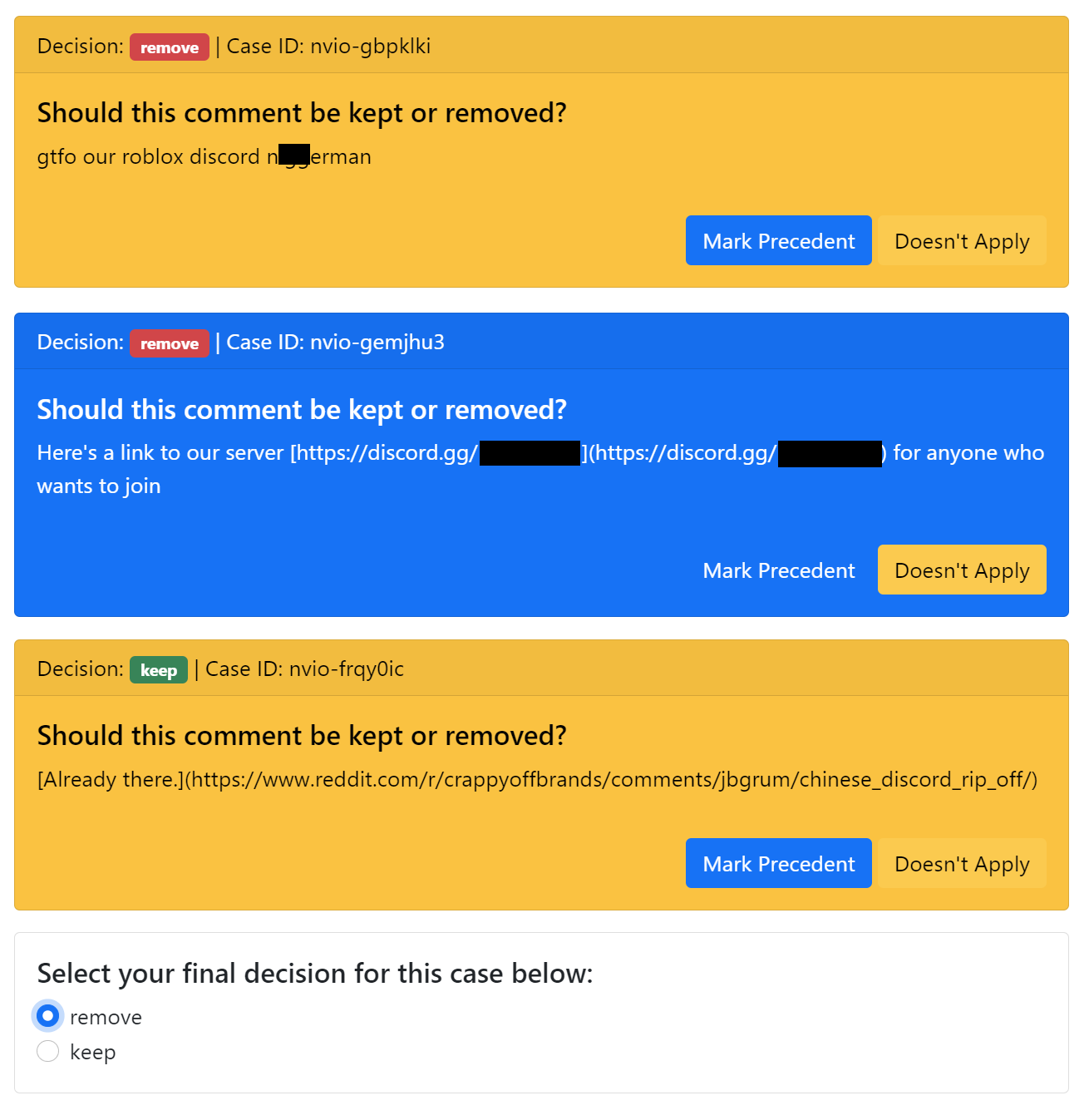}
      \caption{Case law grounding interface for the \textsc{case} condition}
      \label{fig:sc:user-interface:precedents}
  \end{subfigure}
  \hfill
  \begin{subfigure}[t]{0.48\linewidth}
      \vskip 0pt
      \includegraphics[width=\linewidth]{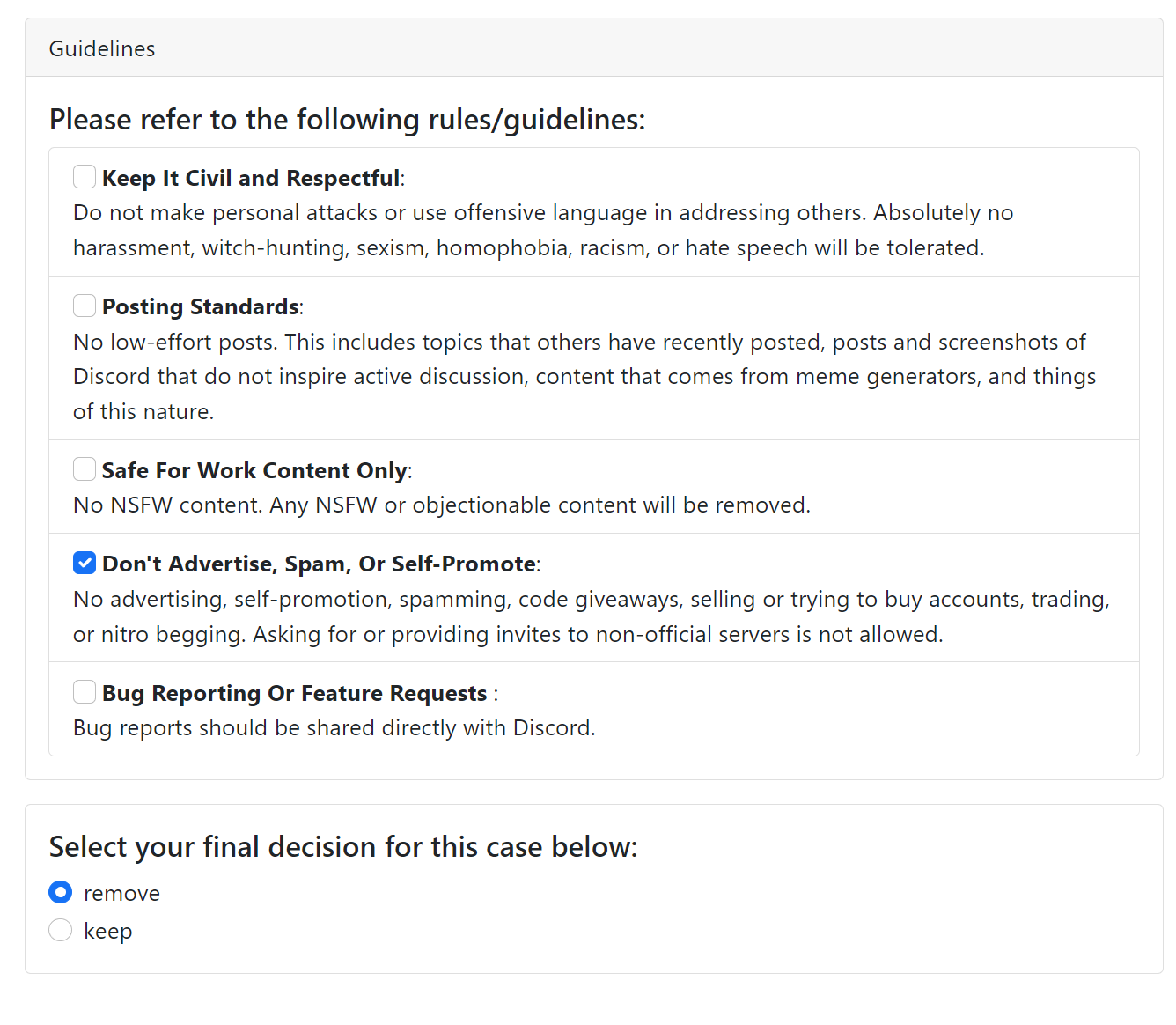}
      \caption{Constitutional grounding interface for the \textsc{rule} condition}
      \label{fig:sc:user-interface:rules}
  \end{subfigure}
  \caption{Screenshots of the annotation tool interface used \chm{to collect annotations in the human-led process}.}
  \label{fig:sc:user-interface:all}
\end{figure*}

\subsubsection{Case Selection Tool}

For the human-led decision-making process, we built an annotation interface that is deployed to our participants in the form of a web application (\autoref{fig:sc:user-interface:all}).
We split the annotation interface down the middle, presenting the general usage instructions and the case being judged on the left (\autoref{fig:sc:user-interface:instructions}), and interface components to support making grounded decisions placed on the right.

We use the right side interface to facilitate the steps for conducting decision-making following the \textit{case law} grounding process.
During annotation, the right side of the interface is populated with a list of cases retrieved through the case retrieval step (\autoref{fig:sc:user-interface:precedents}).
Under each retrieved case, the annotator can click on a button to toggle whether a case should be considered a precedent or not.
Once the annotator is done evaluating each case, they are asked to provide a final decision given their review of the precedents but not constrained to the decisions of them.
While selecting the set of applicable precedents alone is sufficient for the CLG workflow, this last question is included to allow us to evaluate whether enforcing binding precedents (\autoref{sec:design:step:3:deriving-judgments}) is important to final accuracy or not.

Additionally, for our experiments, we adapted the same interface to also facilitate decision-making following a more traditional \textit{constitutional} grounding process.
In the constitutional grounding condition, we use the same general interface but render a list of rules and guidelines associated with each case in our task domain (\autoref{fig:sc:user-interface:rules}) instead of retrieved precedents. 
We also include checkboxes for each rule, prompting the human annotator to indicate any that they applied during their decision-making process.
As we can't derive the final decision directly in constitutional grounding, we rely on the annotators to provide their own final decision, collecting this decision using the same design as the case law grounding version.

\subsubsection{LLM Prompts for Precedent Selection}

To evaluate an automated process where LLMs contribute to decision-making instead of human annotators, we also implemented a series of prompts for LLMs to follow a grounding process similar to the human-led version (\autoref{appendix:prompts}).
In the prompting version of the workflow, we set up the system prompt in a similar way to the instructions section in the annotation interface, providing the instructions for the decision task itself (whether the model is assessing case precedents or constitutional guidelines) as well as task domain specific instructions.
We then provide the case being judged along with the grounding support as part of the user prompt.
Each user prompt assessment was prompted independently, with no conversation history persisted between two different user prompt queries.

For the LLM-prompting version of CLG, we use a case selection prompt (\autoref{appendix:prompts:case}) to evaluate each retrieved case, substituting \texttt{\{input\}} with the case being judged, \texttt{\{precedent\}} with one retrieved past case, and \texttt{\{decision\}} with its past decision.
This prompt was then repeatedly invoked for each retrieved case, producing a binary output for whether the case should be considered a precedent or not.
Following the CLG process, these precedent determinations were then used to derive the final decision.
Unlike the human annotation interface where annotators also produced a final decision based on their review of precedents but were unconstrained by the decisions of those precedents, we could not create a prompt that provided all precedents to the model to produce a single final decision in a way that fit within the context window of the model; thus, non-binding precedents were only evaluated for the human-led process and not the LLM prompting version.
For the LLM-prompting version of constitutional grounding, we use a guideline assessment prompt (\autoref{appendix:prompts:rule}) to evaluate each retrieved case, substituting \texttt{\{input\}} with the case being judged and \texttt{\{instructions\}} with a list of rules and guidelines for making the decision.
We then collected the resulting output directly as the model's final decision.

\section{Experiments}

To evaluate CLG, we set out to answer the following research questions:
\begin{itemize}
    \item \textbf{RQ1}: Does the \textit{case law} grounding approach result in more accurately aligned decisions compared to traditional rule-based \textit{constitutional} grounding?
    \item \textbf{RQ2}: How do specific choices in the \textit{retrieval setup} and \textit{precedent application} affect the alignment accuracy of decisions produced?
    \begin{itemize}
        \item\textbf{RQ2a}: How does the alignment accuracy scale with different retrieval window size ($k$) for precedents?
        \item  \textbf{RQ2b}: How do \textit{binding precedents} affect the alignment accuracy of the decisions?
    \end{itemize}
\end{itemize}
 
\begin{figure}[t]
  \centering
  \begin{subfigure}[t]{0.37\linewidth}
    \centering
    \includegraphics[width=\linewidth]{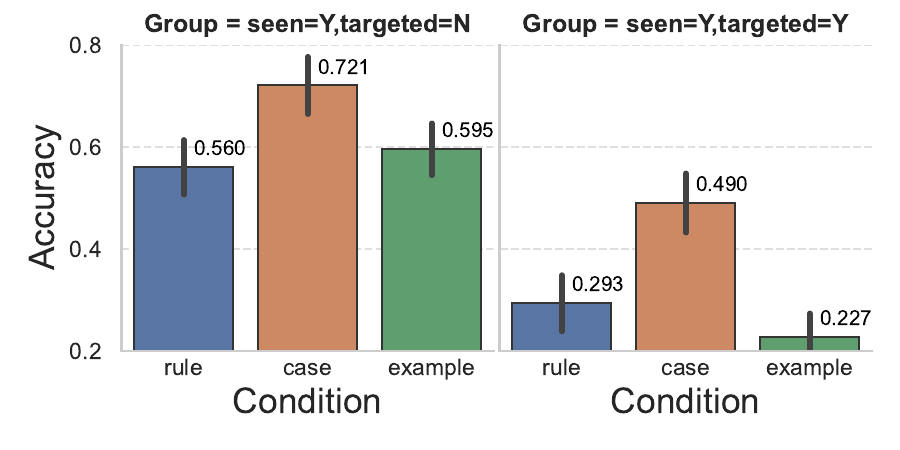}
    \caption{Accuracy of human decisions the \texttt{tox} domain}
  \end{subfigure}~
  \begin{subfigure}[t]{0.57\linewidth}
    \centering
    \includegraphics[width=\linewidth]{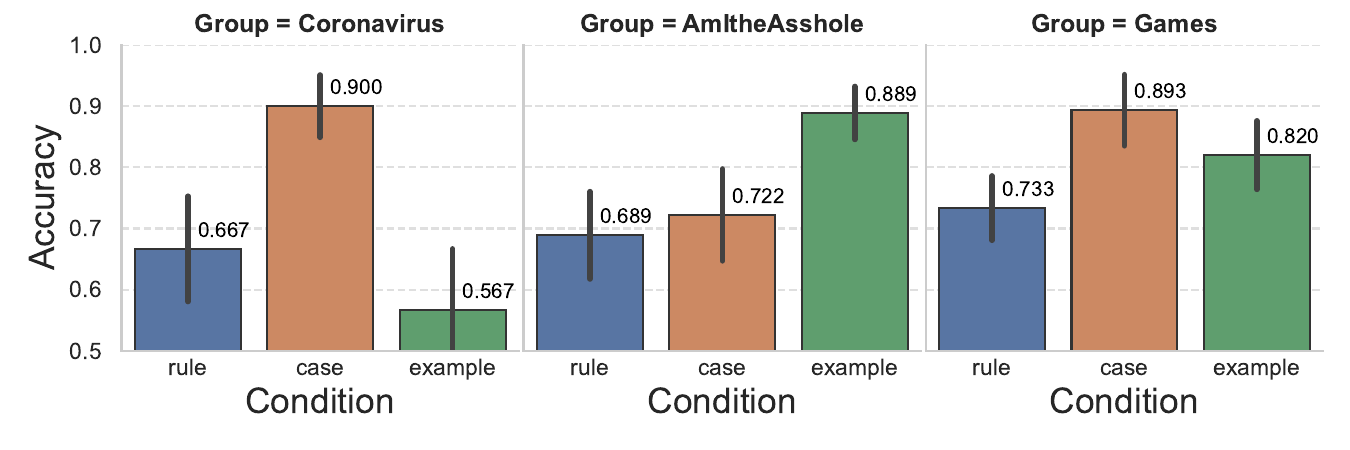}
    \caption{Accuracy of human decisions the \texttt{mod} domain}
  \end{subfigure}
  \caption{\chm{Accuracy of the decisions produced by humans using each grounding approach, \textsc{rule} for constitutional grounding and \textsc{case} for Case Law grounding. For the \textsc{case} condition, results reflect the retrieval setup used during annotation ($k = 15$). We also include an \textsc{example} condition, which presents the accuracy when binding precedents are not enforced and retrieved cases primarily act as \textit{examples} (details discussed in \autoref{sec:results:rq2b}).}}
  \label{fig:results:rq1:human}
  \Description{Bar plots comparing the accuracy of decisions produced by humans using each grounding approach.}
\end{figure}

\subsection{Experiment Setup}

To answer RQ1, we set up two main conditions \textsc{rule}---representing a constitutional grounding process, and \textsc{case}---representing CLG.
We conducted this comparison across 2 types of decision-making tasks (\textit{mod}eration and \textit{tox}icity rating) spanning 5 community / group specific decision-making task instances. In each condition, we evaluated both the \texttt{human}-led and \texttt{llm}-prompting versions of CLG.

For RQ2, we dive into how the choices in the CLG configuration contributed to performance by conducting ablation-style experiments. Specifically, we examined the effects of a smaller retrieval window size on decision accuracy (RQ2a) and we compared the decisions derived from binding precedents with human-produced decisions (non-binding) to examine the need for binding precedents (RQ2b).

\subsubsection{Tasks and Datasets}
\label{sec:experiment:setup:tasks}

We picked two real-world decision-making tasks that involve subjective decisions with social context specific to groups: a comment \texttt{mod}eration task where the decision is to ``keep'' or ``remove'' a comment posted in a Reddit community, and a \texttt{tox}icity rating task where the decision is the scalar rating of the toxicity level of a social media post (on a scale from 1--5).
We selected datasets for ground truth annotations that covered different groups / communities that we could readily recruit for our annotation study. 
This dataset separates comments by the ``subreddit'' it was posted in, which we use to partition distinct communities that will likely have distinct preferences around content moderation.
For the \texttt{tox} domain, we used the toxicity perspectives dataset~\cite{kumar2021designing} which contains a wide range of posts collected from social media platforms that were then evaluated by crowd annotators and rated for toxicity.
This dataset contains disaggregated ratings that are associated with demographic information of the annotator, which we use to segment a general crowd worker population into distinct salient groups.

Within each dataset, we performed additional steps to select for communities and groups with sufficient data size that we could readily recruit that we document in \autoref{appendix:experiments:datasets}. We also randomly shuffled and sub-sampled data for each task domain and group instance (10\% as precedents, 10\% used in both human and LLM decision evaluation). We also divided sampled cases into ``batches'' of 10 each, with each annotator only judging one such batch. In the constitutional grounding condition (\textsc{rule}) we drew from existing rules and guidelines associated with each task domain and group (see \autoref{appendix:experiments:datasets} for details).

\begin{figure}[t]
  \centering
  \begin{subfigure}[t]{0.37\linewidth}
    \centering
    \includegraphics[width=\linewidth]{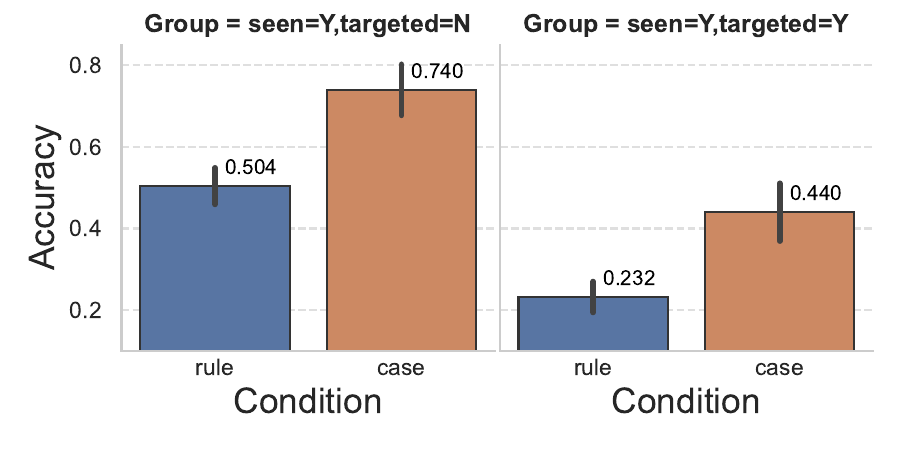}
    \caption{Accuracy of LLM decisions the \texttt{tox} domain}
  \end{subfigure}~
  \begin{subfigure}[t]{0.57\linewidth}
    \centering
    \includegraphics[width=\linewidth]{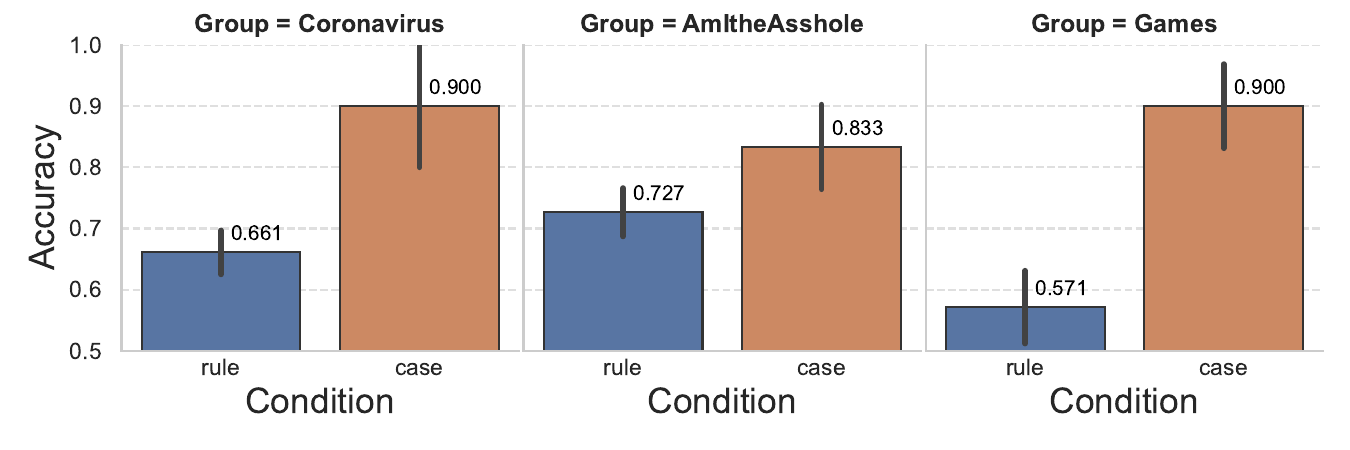}
    \caption{Accuracy of LLM decisions the \texttt{mod} domain}
  \end{subfigure}
  \caption{\chm{Accuracy of the decisions produced by LLMs prompted using each grounding approach, \textsc{rule} for constitutional grounding and \textsc{case} for Case Law grounding. To match human decision-makers, $k = 15$ was also used as the retrieval window for presenting precedents to the LLM in the \textsc{case} condition.}}
  \label{fig:results:rq1:llm}
  \Description{Bar plots comparing the accuracy of decisions produced by LLMs using each grounding approach.}
\end{figure}

\subsubsection{Communities and Groups}
\label{sec:experiment:setup:groups}

A major challenge in making decisions over subjective and socially-defined concepts lies in the locality of preferences---criteria for the same decision task like toxicity rating or moderation can differ considerably between groups and communities~\cite{Koshy2023MeasuringUA}.
To capture this, we further split each task domain into specific task instances defined by the group or community the decisions should align to.

For the \texttt{mod} domain, we selected 3 large communities from Reddit that were captured by the NormVio dataset. For the \texttt{tox} domain, demographic backgrounds of raters can contribute to significant preference differences, so we take the most significant two (prior exposure and observers versus victims) such demographic dimensions identified by prior work~\cite{kumar2021designing}. 
Overall, this resulted in 5 group-defined task instances, details of which we document in \autoref{appendix:experiments:groups}.

\subsubsection{Participant Recruitment and Crowdsourced Annotation}

We recruited participants for the \texttt{mod} tasks directly from the respective subreddits.
For \texttt{tox}, we recruited participants on Prolific, a crowdsourced survey platform. 
In both domains, we used a screening survey to identify participants that would be a good fit.
Human participants in both conditions were paid a base rate of \$15 for completing our annotation study (30-45 minutes), with Prolific participants being paid through the platform and Redditors being paid in the form of an Amazon Gift Card. 
In total, we were able to recruit $n=60$ participants from Prolific and $n=35$ participants from Reddit.
We document the additional details around recruitment in \autoref{appendix:experiments:annotation}.

\subsubsection{Additional Setup Details}
To facilitate evaluations such as the simulation study or examination of binding-ness of precedents, we made additional configurations to the annotation task setup. These details, as well as general configuration parameters, are documented in \autoref{appendix:experiments:additional}.

\subsection{\chm{RQ1: Comparing Case Law and Constitutional Grounding}}
\label{sec:results:rq1}

To answer RQ1, we compare the accuracy at which each grounding approach reproduces the ground truth decisions observed for each group and task. 
For the human-led process, we follow the workflow in \autoref{sec:design:workflows}, using human annotators for each task and group to produce decisions grounded by either constitutional grounding (\textsc{rule}) or case law grounding (\textsc{case}). 
We found that for the human-led process (\autoref{fig:results:rq1:human}), case law grounding was able to produce statistically significantly higher accuracy in 4 out of 5 groups: +16.7 p.p.\footnote{percentage points} for \texttt{seen=Y,targeted=N} in \texttt{tox} (paired t-test, $p = 0.010 < 0.05$), +19.1 p.p. for \texttt{seen=Y,targeted=Y} in \texttt{tox} (paired t-test, $p = 0.019 < 0.05$), +23.3 p.p. for \texttt{Coronavirus} in \texttt{mod} (paired t-test, $p = 0.045 < 0.05$), and +16.0 p.p. for \texttt{Games} in \texttt{mod} (paired t-test, $p = 0.029 < 0.05$).
However, we did not observe any significant difference between the two grounding approaches in the \texttt{AmItheAsshole} in \texttt{mod}, with a difference of 3.3 p.p. (paired t-test, $p = 0.721 > 0.05$).

For the LLM-prompting process, we also follow the prompts in \autoref{sec:design:workflows}.
\chm{We found that for the LLM process (\autoref{fig:results:rq1:llm}), case law grounding was able to produce statistically significantly higher accuracy in 4 out of 5 groups: +23.6 p.p. for \texttt{seen=Y,targeted=N} in \texttt{tox} (paired t-test, $p = 0.004 < 0.05$), +20.8 p.p. for \texttt{seen=Y,targeted=Y} in \texttt{tox} (paired t-test, $p = 0.006 < 0.05$), +23.9 p.p. for \texttt{Coronavirus} in \texttt{mod} (paired t-test, $p = 0.011 < 0.05$), and +32.9 p.p. for \texttt{Games} in \texttt{mod} (paired t-test, $p = 0.006 < 0.05$).}
However, similar to the human-led version, we did not observe any significant difference between the two grounding approaches in the \texttt{AmItheAsshole} in \texttt{mod}, with a difference of +10.6 p.p. (paired t-test, $p = 0.229 > 0.05$).

Overall, these results suggest that for the vast majority of the tested groups and tasks, CLG resulted in decisions that had statistically significantly higher accuracy compared to constitutional grounding.
We find \texttt{AmItheAsshole} in \texttt{mod} to be an outlier---while average accuracy was higher, the difference was not significant, indicating that grounding with cases can be less effective in some situations.

\subsection{\chm{RQ2: Assessing Retrieval and Decision Derivation Configurations}}
\label{sec:results:rq2}

For our second RQ, we examine two configuration aspects of the CLG workflow that may contribute to the final accuracy: the retrieval window size ($k$), and whether to enforce binding precedents (\textsc{case}) as is default in CLG or not (\textsc{example}).

\subsubsection{\chm{RQ2a: Alignment Accuracy and Retrieval Window Size}}
\label{sec:results:rq2a}

\begin{figure}[t]
  \centering
  \begin{subfigure}[t]{0.43\linewidth}
    \centering
    \includegraphics[width=\linewidth]{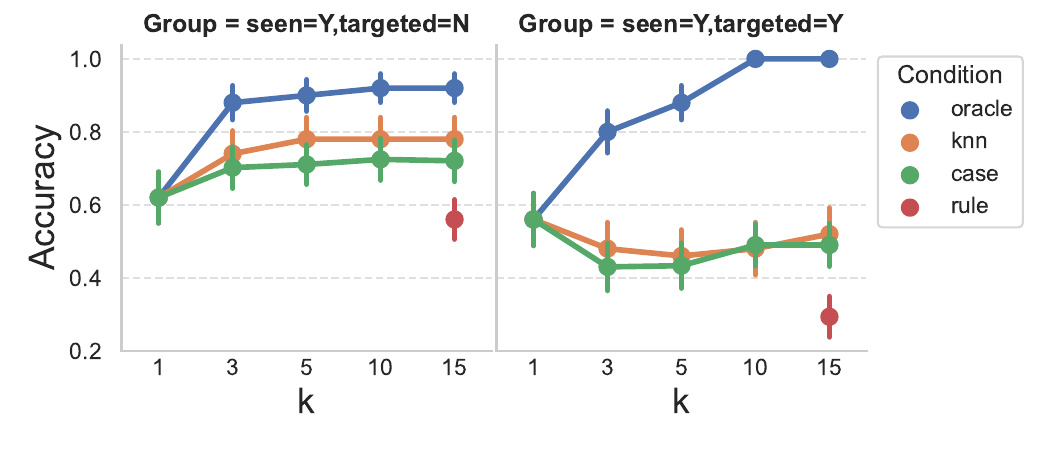}
    \caption{Accuracy scaling on the \texttt{tox} domain for humans}
  \end{subfigure}~
  \begin{subfigure}[t]{0.56\linewidth}
    \centering
    \includegraphics[width=\linewidth]{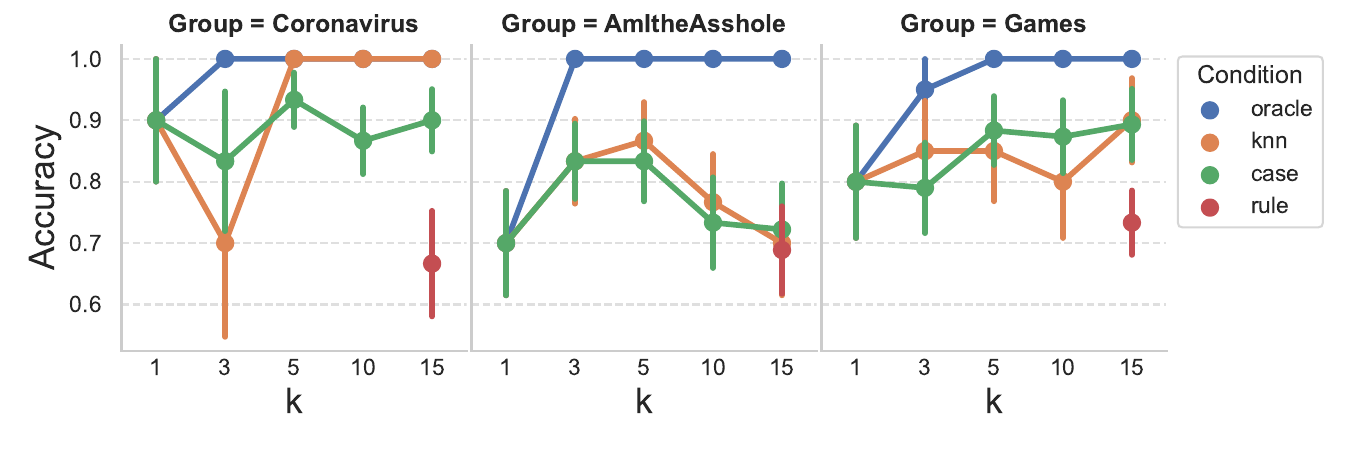}
    \caption{Accuracy scaling on the \texttt{mod} domain for humans}
  \end{subfigure}
  \caption{\chm{Scaling of human decision accuracy given different retrieval window sizes $k \in \{1, 3, 5, 10, 15\}$ for Case Law grounding (simulated results when $k \neq 15$). We also plot two reference lines: \textsc{oracle} indicates the upper-bound accuracy when picking the best available precedent(s) within the retrieval window; \textsc{knn} indicates the accuracy of the consensus decision for each retrieval window. Additionally, we note that \textsc{rule} does not involve retrieval and thus has no window size ($k$), however we plot it at $k = 15$ to match the non-simulated window size $k$ when collecting results for \textsc{case}. }}
  \label{fig:results:rq2a:human}
  \Description{Line plots}
\end{figure}

\begin{figure}[t]
  \centering
  \begin{subfigure}[t]{0.43\linewidth}
    \centering
    \includegraphics[width=\linewidth]{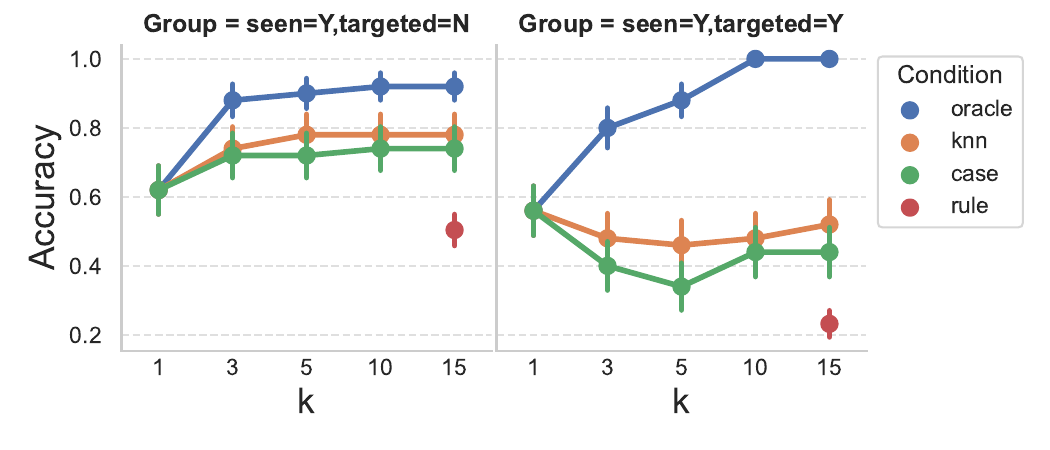}
    \caption{Accuracy scaling on the \texttt{tox} domain for LLMs}
  \end{subfigure}~
  \begin{subfigure}[t]{0.56\linewidth}
    \centering
    \includegraphics[width=\linewidth]{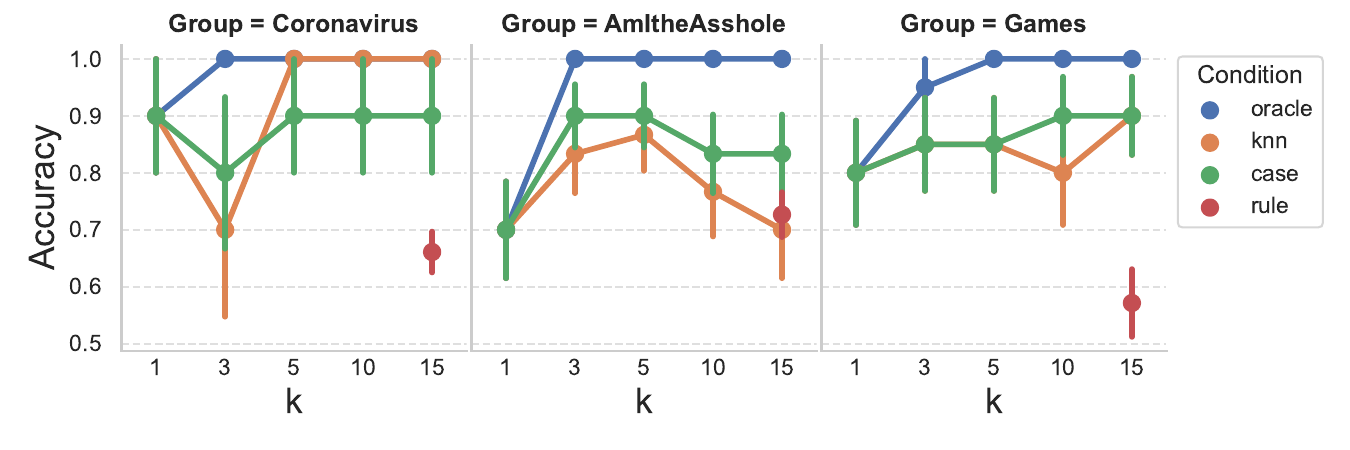}
    \caption{Accuracy scaling on the \texttt{mod} domain for LLMs}
  \end{subfigure}
  \caption{\chm{Scaling of LLM decision accuracy given different retrieval window sizes $k \in \{1, 3, 5, 10, 15\}$ for Case Law grounding. Reference lines for \textsc{oracle} and \textsc{knn} same as in \autoref{fig:results:rq2a:human}. \textsc{rule} indicates outputs for LLM prompts with constitutional grounding, also plotted at $k = 15$.}}
  \label{fig:results:rq2a:llm}
  \Description{Line plots}
\end{figure}

In \autoref{fig:results:rq2a:human} and \autoref{fig:results:rq2a:llm}, we plot the simulated accuracy (\autoref{appendix:experiments:additional}) of CLG (\textsc{case}) in the human-led and LLM-prompting based processes at smaller retrieval window sizes than we used for RQ1 to evaluate whether smaller retrieval windows may be sufficient.
In addition to plotting the scaling of \textsc{case} at different values of $k$ and presenting the \textsc{rule} condition for comparison at $k = 15$, we also plot how 2 additional reference lines behave at different values of $k$: \textsc{oracle} and \textsc{knn}.
The \textsc{oracle} reference line indicates an upper bound on accuracy given a certain retrieval window. For any case being judged, if the retrieval window contains a past case matching ground truth, the \textsc{oracle} will have perfect accuracy.
We observe that at small retrieval windows, it is often the case that across the entire set of judged cases, even the \textsc{oracle} cannot achieve perfect accuracy, as for some cases no correct selection of precedents exists. However, as $k$ becomes larger, we see oracle more consistently achieving perfect accuracy.
The \textsc{knn} reference line represents a strategy that always selects all precedents and relies on the tie-breaking mechanism in CLG to produce the final decision. 

In both human and LLM processes, we did not find any consistent correlation between $k$ values and accuracy across the tested groups and domains, nor did we find statistically significant differences between neighboring $k$ values when it came to final accuracy, suggesting that for many of the individual groups and domains tested, smaller $k$ values may be sufficient to achieve a similar level of accuracy.

However, looking across groups we also observed the following behavior patterns that we believe suggest against picking very small $k$ values to reduce decision cost, especially since it is not possible to evaluate scaling in practice when ground truth is not available:

\textbf{Performance plateau behavior.}
In most of the groups and domains (excluding the outlier \texttt{AmItheAsshole} from before), accuracy tends to reach a plateau where further scaling of $k$ contributes little to accuracy. 
However, the speed at plateauing varies by the group and domain: 
For example, groups like \texttt{seen=Y,targeted=N} or \texttt{Coronavirus} seem to reach a plateau at $k = 3$ and $k = 5$, while \texttt{Games} still sees accuracy improvements at $k = 10$.
While this behavior is not too surprising by itself---with more dissimilar cases retrieved, we should expect additional cases to be less and less relevant to the final decision---there does not seem to be a consistent $k$ that ensures plateauing across groups and tasks.
We hypothesize that this variation across groups is likely to do with aspects like the ``density'' (average distance) of past cases that can be retrieved and distributional properties of past cases around the judged cases.

\textbf{Unreliable for retrieval windows that are too small.}
We also observe that very small values of $k$ such as $k = 1$ or $k = 3$ result in unpredictable performance.
For \texttt{seen=Y,targeted=Y} and \texttt{Coronavirus}, we even saw drops in performance at lower $k$ values.
While these drops are not significant when viewed across all judged cases, they indicate that accuracy is more sensitive to $k$ when $k$ is small---moving between retrieval thresholds can impact accuracy on some portions of cases.
We hypothesize that this may be due to case-based approaches naturally being more sensitive when the retrieval window is low, as small changes could result in relevant precedents not making it past the threshold or immediate local neighborhoods primarily comprising of similar but less relevant cases.

Ultimately, we don't find generalizable correlations between changes in the retrieval window $k$ and final performance, only noting that for any task / group, $k$ should be sufficiently large so as to reach the performance plateau; otherwise (when $k$ is too small) performance can suffer.

\subsubsection{RQ2b: Binding Precedents versus Cases as Examples}
\label{sec:results:rq2b}

In prior work around grounding human decision-making in crowdsourcing settings, \textit{examples}~\cite{Wu2017ConfusingTC} have been noted as an important feature for constructing more effective guidelines or ``constitutions''.
Unlike CLG, however, constitutional grounding examples are selected and incorporated during the construction of the constitution, meaning that they do not adapt to properties of the case being judged.
This raises the question of whether simply retrieving and presenting prior related cases for a new case in the form of \textit{examples} alone is sufficient for improving grounding accuracy or if the \textit{binding} nature of precedents---where decisions must derive directly from selected precedent sets---needs to be applied for CLG to produce accurately grounded decisions.

To examine this, during the human annotation process, we not only collected participants' judgments on whether each precedent \textit{applied}, we also asked participants to give a final decision on the judged case after reviewing the list of retrieved prior cases without being constrained to the consensus of the cases they selected as precedents (as noted in \autoref{sec:design:workflows}).
In \autoref{fig:results:rq1:human}, we include an additional condition---\textsc{example}---to reflect the accuracy of this configuration.
Comparing the accuracy produced by non-binding precedents (\textsc{example}) to that of constitutional grounding (\textsc{rule}), we note the absence of statistically significant differences (paired t-test, $p = 0.635, 0.360, 0.458, 0.266 > 0.05$ for \texttt{seen=Y,targeted=N}, \texttt{seen=Y,targeted=Y}, \texttt{Coronavirus}, \texttt{Games}) in most of the groups, despite the fact that we did observe statistically significant improvements to accuracy when using binding precedents (\textsc{case}).
Here, again, \texttt{AmItheAsshole} is an outlier, with \textsc{example} showing a statistically significant improvement over \textsc{rule} (+20.0 p.p, $p = 0.020 < 0.05$) even though binding precedents (\textsc{case}) did not show significant improvements.

\section{Discussion}

\subsection{Limitations of Our Implementation}

In this work, we explored a relatively simple implementation of case law for grounding decisions where reasoning around cases is abstracted simply into a selection process.
However, in the legal space, there are many specific conventions around how precedents are actually used~\cite{ashley1989modelling,Sunstein2021AnalogicalR}.
While the degree to which case-based and analogical reasoning can be applied is still openly debated in the legal realm~\cite{Caputo2024AlignmentAJ}, there are opportunities to improve the value of grounding cases by making use of full-fledged analogical reasoning.

Furthermore, we also observed differences between groups / communities in how CLG affected decision alignment accuracy in our experiments, with notable cases like \texttt{AmItheAsshole} seeing a very different performance pattern.
While we don't have sufficient results to examine this more comprehensively, we do hypothesize that the permissiveness and concreteness of the community's internalized criteria itself may be at play. 
Compared to other communities examined, \texttt{AITA} was much more permissive about what content is acceptable: even in ambiguous cases (like suspected fabricated stories), enforcement of moderation tended to lean lenient. 
This inherent acceptance of ambiguity could contribute to less visible differences between constitutional and case law style grounding, which specifically focuses on challenges with ambiguous concepts.

\subsection{Integrating CLG into Decision-Making Ecosystems}

While we primarily compared CLG and constitutional grounding in this work, both approaches offer unique values in the broader ecosystem of decision-making.
When we look at prior examinations of community governance~\cite{hwang2022rulemakingwiki}, we often see a combination of rules and cases being used throughout the process---with cases contributing to ``legislative'' processes that produce the very rules and guidelines that are adopted while also being applied in debates over ambiguous cases.
Similar patterns of adopting both abstractions (constitutional rules) and specifics (case-based analogical reasoning) are common in high-stakes decision-making tasks such as with jurisprudence~\cite{Caputo2024AlignmentAJ}.

Ultimately, constitutional and case-law grounding are suited to different challenges in decision-making, and with this work, our goal is to bring to light the need for more case-based approaches around the challenging decision-making tasks like those involved in online community governance or in areas like AI alignment. 
We believe there is much to be explored around future processes that combine both forms of grounding to efficiently produce high-quality decisions---both in human-led and AI automated settings.
\section{Conclusion}

In this paper, we present \textit{case law grounding} (CLG), a novel workflow that uses precedents for aligning decisions to socially constructed norms and preferences of groups and communities.
We evaluate two versions of CLG, a human-led process and an automated LLM process, on 2 decision-making tasks spanning 5 instances. We also examine the impacts of different configurations of CLG such as retrieval window size and binding or non-binding precedents. Finally, we discuss the limitations of CLG and how CLG could fit into the broader ecosystem around grounding decision-making in sociotechnical systems.


\bibliographystyle{ACM-Reference-Format}
\bibliography{bibliography}

\appendix

\section{Experiment Setup Details}

\subsection{Dataset Details}
\label{appendix:experiments:datasets}
\paragraph{Preprocessing and Selection for Moderation (mod)}
From this dataset, we selected the top 10 subreddits with a sufficiently large size of examples ($n > 100$) and did not exhibit significant skew towards either decisions of ``keep'' or ``remove''.
We then further narrowed down this set based on avoiding multiple communities around similar topics and prioritizing communities that were easier to recruit from, ultimately sub-sampling this dataset down to 3 communities: \texttt{r/Coronavirus}, \texttt{r/AmItheAsshole}, \texttt{r/Gaming}.
Ground truth labels for each case are established by the actual moderation decision applied to each comment.

\paragraph{Preprocessing and Selection for Toxicity (tox)}
Based on the findings from the original paper, we sampled from 3 disjoint intersectional demographic slices (see \autoref{sec:experiment:setup:groups}) that were indicated to have the most significant differences with each other: \texttt{seen=N,targeted=N}, \texttt{seen=Y,targeted=N}, \texttt{seen=Y,targeted=Y}. Ground truth labels for each case in each group are established by taking the mean rating for that group rounded to the nearest integer.

\paragraph{Constitutional Grounding Material}
When evaluating the constitutional grounding condition (\textsc{rule}), we use published community guidelines for each subreddit as approximations of a high-quality constitution for groups in \texttt{mod}. Meanwhile, for the \texttt{tox} domain, we adapt the annotation guidelines given to annotators for rating toxicity in the original dataset paper into a list of rules that served as the constitution for all groups.
To address the fact that community guidelines may have changed over time since being first collected, we checked the most current community guidelines for each of the 3 groups (subreddits) in the \texttt{mod} domain using the Wayback Machine\footnote{\url{https://web.archive.org/}} to confirm that no significant changes (such as adding or removing rules) were made to the guidelines across the time span of cases evaluated.

\subsection{Communities and Groups}
\label{appendix:experiments:groups}

\paragraph{Moderation (mod)}

For this domain, we ended up selecting the following 3 subreddits:
\begin{itemize}
    \item \texttt{r/Games}---a broad interest group around game content and the games industry
    \item \texttt{r/Coronavirus}---a community focused around high-quality science-informed discussions related to COVID-19
    \item \texttt{r/AmItheAsshole}---a collective space for sharing and opining over a variety of interpersonal conflict scenarios
\end{itemize}

\paragraph{Toxicity (tox)}

While raters do not necessarily form a ``community'', we can still view \textit{demographic features} as a way to separate the population into groups that represent distinct norms and preferences.
In our case, we stratify the general population by taking two demographic properties identified to be significant correlating factors~\cite{kumar2021designing} of toxicity ratings: whether a rater has often \texttt{seen} toxic content when interacting with social media (negatively correlated with scores), and whether the rater has been personally \texttt{targeted} by toxic content (positively correlated with scores).
Out of the 3 sensible combinations---\texttt{seen=Y,targeted=N}, \texttt{seen=N,targeted=N}, and \texttt{seen=Y,targeted=Y}---we found it challenging to find participants for \texttt{seen=N,targeted=N}, so we did not include that group in the evaluations, leaving just 2 combinations.

\subsection{Crowd Recruitment Details}
\label{appendix:experiments:annotation}
As we envision human-led processes for grounding subjective decisions (whether constitutional or case law) to be used in practice by groups and communities themselves rather than by external populations, we set out to recruit participants that would demographically best match the original group or community producing the ground truth data.
To achieve this goal, we applied pre-screening surveys in each of our recruitment efforts (Appendix~\ref{appendix:screener}), with participants first asked to fill out the screening survey, after which we would select and assign conditions to them.

As each of the \texttt{mod} groups represents a specific subreddit community, we recruited participants by running Reddit Ads targeted to subreddits used in our task domains.
Participants were asked to fill out the screening survey and self report subreddits they frequented within a list of communities (subset of the original top-10 candidates \autoref{sec:experiment:setup:tasks}, removing 1 subreddit that had been removed).
Then, participants who reported frequenting at least one of the 3 target subreddits would be allocated a batch and condition, balancing such that each community, condition, and batch had a target of 3 annotators.
At the end, this resulted in annotations on 3 batches for \texttt{r/AmItheAsshole}, 2 for \texttt{r/Games}, and 1 for \texttt{r/Coronavirus}.

In the \texttt{tox} domain, original annotations in the original dataset were provided by Amazon Mechanical Turk (AMT) workers.
However, recent work has shown concerns around AMT~\cite{Marshall2023WhoBA} annotations that have developed since the original dataset was collected.
To avoid issues with introducing significantly more quality control components than the original task, we opted to recruit from Prolific, as it is a similar crowd work platform with (as of our deployment) lower rates of data quality issues.
Participants were screened with two questions about their prior exposure to toxic posts, with these answers then used to assign them to one of 3 population groups (see \ref{sec:experiment:setup:groups}). 
During recruitment, we also found that no crowd workers reported being in the \texttt{seen=N,targeted=N} group during screening. While this was ultimately unsurprising, we opted to drop this group.

As both experiment domains involve reading content that is potentially offensive, we made sure to prominently disclose this aspect during recruitment.
We also allowed participants to drop out at any point if they experienced discomfort with the annotation.
Annotators electing to withdraw consent during the task were pro-rated payment for the amount of annotations completed, with their data excluded from the evaluation.
Additionally, to incentivize participation from Redditors and reduce drop-out between screening and completion, we provided an additional bonus of \$5 to Redditors who completed our annotation task within 24 hours of assignment. 

\subsection{Additional Implementation Details}
\label{appendix:experiments:additional}

\textbf{Language Model and Embedding Model.}
As our goal was not to evaluate a variety of language models, we opted to keep the models used to produce embeddings and the LLM used to execute prompts consistent and fixed across all conditions and task domains.
For the embedding model, we used the general purpose embedding offered by OpenAI\footnote{https://openai.com/blog/new-embedding-models-and-api-updates} (``\texttt{text-embedding-3-large}'' ) with a distance metric based on cosine similarity for ranking relevant cases.
As precedent retrieval only depends on the case being judged, we pre-computed the ordered list of retrieved past cases with $k = 15$ as the retrieval window and provided this for human annotations, with smaller window sizes simulated post-hoc.
For the LLM used during prompting, we used the \texttt{gpt-4-turbo-preview} model offered by OpenAI. 

\textbf{Retrieval Window Simulations.}
In RQ2a, we evaluate how alignment accuracy may be affected by the retrieval window size, specifically examining different options of $k \in \{1, 3, 5, 10, 15\}$.
However, directly evaluating different retrieval outputs in the human-led workflow would be very costly, especially given the observation that the set of retrieved examples at smaller $k$ values would just be subsets of those at larger $k$.
Thus, instead of separately testing different values of $k$, we only have the human participants and LLM select precedents over the largest retrieval window size $k = 15$.
We then simulate the outcome of precedent selection for smaller $k$ by dropping any selected precedents that would fall outside the retrieval window at that size.

\textbf{Consensus Tie-Breaking for Binding Precedents.}
When deriving final decisions based on precedents selected, it is possible for the collection of precedents selected by either humans or LLMs to contain a plurality of decision outcomes.
While in the legal realm such conflicts would indicate further reasoning and deliberation, in CLG we apply a simpler deterministic approach to resolve conflicts based on \textit{approval voting}.
We collect the past decision of each selected precedent, and then take the \textit{mode} (most common value) to be the final decision. 
In the event of a tie where there are multiple values for the mode, we break the tie by selecting the decision that has a corresponding precedent case that is \textit{closest} (retrieval distance) to the original example being judged.
Should the set of precedents be empty, we treat this case as a tie between all retrieved cases with each getting no votes.

\section{LLM Prompts}
\label{appendix:prompts}

\subsection{Task System Prompts}
\label{appendix:prompts:shared}
System prompt for \textsc{mod} \textsc{case}:
\begin{verbatim}
You are an AI assistant that is assisting a moderator analyze 
comments posted in an online community. 

For each comment you are analyzing, you will be shown past 
comments that the moderator has either removed or kept.

Your task is to decide whether the decision on each of the 
past comments would be relevant to the new comment.

A relevant decision is a decision where the new comment 
exhibits some similarities to the past comment.
Think step by step.
\end{verbatim}

System prompt for \textsc{mod} \textsc{rule}:
\begin{verbatim}
You are an AI assistant that is assisting a moderator analyze 
comments posted in an online community. 

The moderator will provide a list of rules that posts and 
comments in the community must adhere to. 

Based on the list of rules, you should analyze the comment, 
and respond with either "keep" or "remove".
Think step by step.
\end{verbatim}

System prompt for \textsc{toxicity} \textsc{case}:
\begin{verbatim}
You are an AI assistant that is assisting a user in organizing
their online feed by rating whether content is toxic or not. 

For each post in the feed, you will be shown past posts that 
the user has rated toxicity for on a scale from 1 to 5.

Your task is to decide whether the toxicity rating of the 
past post would be relevant to the new post.

A relevant rating is a rating where the new post exhibits 
some similarities to the past post.
Think step by step.
\end{verbatim}

System prompt for \textsc{toxicity} \textsc{rule}:
\begin{verbatim}
You are an AI assistant that is assisting a user in 
organizing their online feed by identifying whether content 
is toxic or not.

The user will provide a list of indicators of what content is 
commonly toxic.

Based on the list, you should analyze the content, and 
respond with a number between 1 and 5, with 1 indicating 
the content is not toxic at all, and 5 indicating that the
content is extremely toxic.
Think step by step.
\end{verbatim}

\subsection{Prompts for Case Law Grounding}
\label{appendix:prompts:case}
Prompt for case selection:
\begin{verbatim}
The comment you are analyzing is:
{input}

The moderator has judged the following 
similar comment in the past:
{precedent}

The moderator made a decision of "{decision}".
Answer with "relevant" if the past comment should 
be considered in the decision for the new comment, or 
"not relevant" if it should not.
\end{verbatim}

\subsection{Prompts for Constitutional Grounding}
\label{appendix:prompts:rule}
Prompt for rule determination:
\begin{verbatim}
Below are a list of guidelines for making your decision:

{instructions}

The content you are deciding is:
{input}

Assess whether each rule applies, and then make your
final decision.
Think step by step.
\end{verbatim}

\section{Screening Survey}
\label{appendix:screener}

\subsection{Screening Questions for \textsc{toxicity}}
These questions are adapted from a subset of questions taken from survey questions annotators had to answer in the original toxicity perspectives dataset~\cite{kumar2021designing}. 

\begin{enumerate}
    \item What types of sites do you use regularly?
    \begin{itemize}
        \item Social Networking (Facebook, Twitter)
        \item Video (YouTube, Twitch)
        \item News (CNN, Fox, NYT, WSJ)
        \item Community Forums (Reddit, Craigslist, 4chan)
        \item Email or messaging (Gmail, WhatsApp, Facebook Chat)
    \end{itemize}
    \item For the sites you use, have you ever seen posts or comments you consider toxic?
    \begin{itemize}
        \item Yes
        \item No
    \end{itemize}
    \item Have you ever personally been the target of toxic posts or comments?
    \begin{itemize}
        \item Yes
        \item No
    \end{itemize}
\end{enumerate}

Questions 2 and 3 were used to determine which group the annotator was placed in.

\subsection{Screening Questions for \textsc{mod}}
The subreddits used in these questions reflect the top-10 subreddits selected during the process described in \autoref{sec:experiment:setup:tasks}. Participants were only recruited for the task if they indicated that they frequented one of the 3 communities selected in \ref{sec:experiment:setup:groups}.

\begin{enumerate}
    \item Do you frequent any of the following subreddits? Check all that apply.
    \begin{itemize}
        \item \texttt{r/Coronavirus}
        \item \texttt{r/classicwow}
        \item \texttt{r/Games}
        \item \texttt{r/heroesofthestorm}
        \item \texttt{r/ShingekiNoKyoujin}
        \item \texttt{r/AmItheAsshole}
        \item \texttt{r/CanadaPolitics}
        \item \texttt{r/RPClipsGTA}
        \item \texttt{r/LabourUK}
    \end{itemize}
    \item Are you a moderator in any of the aforementioned subreddits?
    \begin{itemize}
        \item \texttt{r/Coronavirus}
        \item \texttt{r/classicwow}
        \item \texttt{r/Games}
        \item \texttt{r/heroesofthestorm}
        \item \texttt{r/ShingekiNoKyoujin}
        \item \texttt{r/AmItheAsshole}
        \item \texttt{r/CanadaPolitics}
        \item \texttt{r/RPClipsGTA}
        \item \texttt{r/LabourUK}
    \end{itemize}
\end{enumerate}

\end{document}